\documentstyle[12pt]{article}

\newcommand{\lbl}[1]{\label{eq:#1}}
\newcommand{\rf}[1]{(\ref{eq:#1})}
\newcommand{\vs}[1]{\rule[- #1 mm]{0mm}{#1 mm}}
\newskip\humongous \humongous=0pt plus 1000pt minus 1000pt
\def\caja{\mathsurround=0pt}

\newif\ifdtup
\def\panorama{\global\dtuptrue \openup1\jot \caja
        \everycr{\noalign{\ifdtup \global\dtupfalse
        \vskip-\lineskiplimit \vskip\normallineskiplimit
        \else \penalty\interdisplaylinepenalty \fi}}}
\def\eqalignno#1{\panorama \tabskip=\humongous
        \halign to\displaywidth{\hfil$\displaystyle{##}$
        \tabskip=0pt&$\displaystyle{{}##}$\hfil
        \tabskip=\humongous&\llap{$##$}\tabskip=0pt
        \crcr#1\crcr}}
\newcommand{\eq}{\vs{2}\begin{equation}}
\newcommand{\en}{\\[2mm]\end{equation}}
\newcommand{\bea}{\begin{eqnarray}}
\newcommand{\ena}{\end{eqnarray}}
\newcommand{\lapprox}{%
\mathrel{%
\setbox0=\hbox{$<$}
\raise0.6ex\copy0\kern-\wd0
\lower0.65ex\hbox{$\sim$}
}}
\newcommand{\gapprox}{%
\mathrel{%
\setbox0=\hbox{$>$}
\raise0.6ex\copy0\kern-\wd0
\lower0.65ex\hbox{$\sim$}
}}

\newcommand{\NP}[1]{Nucl.\ Phys.\ {\bf #1}}

\newcommand{\PL}[1]{Phys.\ Lett.\ {\bf #1}}

\newcommand{\AN}[1]{Ann. Phys. {\bf #1}}

\newcommand{\PRev}[1]{Phys.\ Rev.\ {\bf #1}}
\newcommand{\PRL}[1]{Phys.\ Rev.\ Lett.\ {\bf #1}}
\newcommand{\MPL}[1]{Mod.\ Phys.\ Lett.\ {\bf #1}}

\begin{document}

\begin{titlepage}

\rightline{IPNO/TH 94-53}

\rightline{hep-ph/9411253}
\vspace*{1cm}
\begin{center}

\indent

{\Large{\bf{ Generalized Chiral Perturbation Theory}}}

\vspace{2cm}

{\bf{M. Knecht}} and {\bf {J. Stern}}

\indent

{\sl Division de Physique Th\'eorique
\footnote{Unit\'e de Recherche des Universit\'es Paris XI et Paris VI
associ\'ee au CNRS.}, Institut de Physique Nucl\'eaire\\
F-91406 Orsay Cedex, France} \\

\indent

\vspace{2cm}

\end{center}

\indent

\begin{abstract}

The Generalized Chiral Perturbation Theory enlarges the framework of the
standard $\chi$PT, relaxing certain assumptions which do not necessarily follow
from QCD or from experiment, and which are crucial for the usual formulation of
the low energy expansion. In this way, the experimental tests of the
foundations of the standard $\chi$PT become possible. Emphasize is put on
physical aspects rather than on formal developements of G$\chi$PT

\end{abstract}

\vspace{5cm}

\noindent
{\it{Contribution to the second edition of the Da$\Phi$ne Physics Handbook, L.
Maiani, G. Pancheri and N. Paver Eds., INFN, Frascati, to appear.}}

\end{titlepage}
\vfill \eject

\clearpage
\setcounter{page}{1}
\renewcommand{\thefootnote}{\fnsymbol{footnote}}

\indent

\pagestyle{plain}
\setcounter{equation}{0}
\setcounter{subsection}{0}
\setcounter{table}{0}
\setcounter{figure}{0}

\section{\bf Introduction}

\indent
Up to now, very little is known about the chiral structure of the QCD ground
state. The fact that in the limit of vanishing quark masses $m_u=m_d=m_s=0$,
the
chiral symmetry of QCD is spontaneously broken down to $SU_V(3)$ just ensures
the existence of 8 Goldstone bosons coupled to 8 conserved axial currents. The
strength of this coupling defines a mass scale
$F_0=F_\pi\vert_{m_u=m_d=m_s=0}\simeq 90$ MeV, which is characteristic of
spontaneous chiral symmetry breaking. Indeed, $F_0$ is a long range order
parameter which plays a fundamental role : a non-zero value of $F_0$ not only
signals a broken symmetry phase but, moreover, it is necessary for the
spontaneous breakdown to occur. The fact that $F_0$ is about ten times smaller
than the typical mass-scale $\Lambda_H\sim 1$ GeV of lightest massive bound
states $(\rho, N,\cdots)$ remains unexplained and it raises a question : What
is
the natural size of other order parameters, such as the quark condensate
(further examples will be given shortly),
\eq
B_0=-F_0^{-2}<\bar uu>_0=-F_0^{-2}<\bar dd>_0=-F_0^{-2}<\bar ss>_0\ ,\lbl{cond}
\en
which describe the chiral structure of the massless QCD vacuum ?

Generalized Chiral Perturbation Theory (G$\chi$PT) is motivated by the
observation that
nothing from our present theoretical or experimental knowledge allows us
to answer
the above question in one way or the other : $B_0$ could be
as large as the bound
state scale $\Lambda_H\sim 1$ GeV or as small as the fundamental order
parameter
$F_0\sim 90$ MeV. The main purpose of G$\chi$PT is to provide a sufficiently
precise and broad theoretical framework which would allow to answer the above
question experimentally.

It is being assumed for a long time that the chiral order in the QCD ground
state
resembles the magnetic order of a ferromagnet : The dominant effect is the
quark
condensate \rf{cond} - the analogue of the spontaneous magnetization. In
particular,
this order parameter dominates the (linear) response of the system to a small
perturbation by quark masses \cite{GOR} - the analogue of an external
magnetic field. This
picture, which underlies the standard $\chi$PT \cite{GL1}, is
certainly appealing
by its simplicity. However, before one takes it for granted,
one
should remember that Nature offers examples of magnetic systems, in which the
broken symmetry phase is characterized by a magnetic order of entirely
different
kind, like antiferromagnets \cite{And}. In this case, the average magnetization
is not
necessarily the relevant order parameter. Depending on structural details, the
spontaneous magnetization of an antiferromagnet can be small or even vanish, to
the extend that the ground state approaches the N\'eel-type magnetic order. The
response of the system to an external magnetic field is then dominated by
different order parameters, and it can become non-linear even for not too
strong
fields.

Phenomena similar to antiferromagnetism could be expected in QCD, if $B_0$
(normalised at the scale $\mu=\Lambda_H$) happened to be as small as 100 MeV
$\sim F_0$ \cite{FSS}. Clearly, even a small but non-vanishing condensate would
still
dominate the response to a perturbation by mathematically small quark masses.
However, contrary to the case of magnetic systems, where the strength of
external fields can be
chosen at will, Nature has already made its choice of the size of quark masses.
In the real world, $m_u,\ m_d$ and $m_s$ are certainly small compared to the
hadronic mass scale $\Lambda_H$. Yet, they need not be small enough to
guarantee
the linear response to the explicit symmetry breaking, which would be dominated
by the condensate $B_0$.

G$\chi$PT \cite{FSS}-\cite{KMS} provides a
systematic extension of the standard $\chi$PT
which allows to incorporate the above possibility into its formalism.

\section {\bf Expansion of Goldstone Boson masses}

\indent
If quark masses are switched on, Goldstone bosons acquire a mass. Since in the
real world $m_u,\ m_d,\ m_s\ll\Lambda_H$, the Goldstone boson masses can be
expanded in powers of quark masses.

The coefficients of this expansion are order parameters defined in terms of
correlation functions of the currents
\eq
S^{a}(x)=\bar\psi(x){\lambda^{a}\over 2}\psi(x)\ ,\ \
P^{a}(x)=\bar\psi(x){\lambda^{a}\over 2}i\gamma_5\psi(x)\ .\lbl{sp}
\en
If, for a given value of $B_0\ne 0$, quark masses are sufficiently small, the
linear term in this expansion dominates. Otherwise, the prominent role is
played
by a different order parameter $A_0$, defined by a two point correlator of the
scalar and pseudoscalar densities \rf{sp} :
\eq
\delta^{ab}\Pi_{SP}(q^2)={i\over F_0^2}
\int dx\ e^{iqx}<0\,\vert T\left\{
S^a(x)S^b(0)-P^a(x)P^b(0)\right\}\vert\, 0>
\lbl{A0}
\en
where $a,b=1\cdots 8$, and $\vert\,0>$ denotes the ground state of the
massless QCD. The
small $q^2$ behaviour of the two point function \rf{A0} is dictated
by chiral Ward identities,
\eq
\Pi_{SP}(q^2)={B^2_0\over q^2}+{5\over 96\pi^2}\ {B_0^2\over F_0^2}
\ln\left({\mu^2\over -q^2}\right) + A_0(\mu) + O(q^2)\ ,\lbl{Pisp}
\en
where the $\mu$ dependence of the parameter $A_0$ compensates the scale
dependence of the logarithm. $A_0$ is indeed an order parameter of chiral
symmetry breaking : The operator in \rf{A0} does not contain the singlet
representation of the chiral symmetry group. If, for a given value of quark
masses, $B_0$ turns out to be sufficiently small (i.e. if the ground state
effectively behaves like an antiferromagnet), the pseudoscalar meson masses
start to be dominated by the order parameter $A_0$.

In order to obtain the expansion of Goldstone boson masses explicitly, it is
sufficient
to consider for $m_q\ne 0$ the two-point function of the divergence of the
axial current at zero momentum transfer, where it is related by the well known
Ward identity to the quark condensate. Separating in this Ward identity the
pseudoscalar meson contribution and expanding the remainder in powers of $m_q$,
one obtains $(m_u=m_d=\widehat m$, for simplicity) :
\bea
F^2_\pi M^2_\pi&=&2{\widehat m}F^2_0B + 4{\widehat
m}^2F^2_0 A_0(\mu)+F^2_\pi\delta M^2_\pi(\mu)\nonumber\\
F^2_K M^2_K&=&(m_s+{\widehat m})F^2_0B+(m_s+{\widehat
m})^2F^2_0 A_0(\mu)+F^2_K \delta M^2_\pi(\mu)\lbl{exp}\\
F^2_\eta M^2_\eta&=&{2\over 3}(2m_s+ {\widehat m})\ F^2_0B+{4\over 3}
(2m^2_s+{\widehat m}^2)
F^2_0A_0(\mu)+
{8\over 3}(m_s-{\widehat m})^2F^2_0Z^P_0+F^2_\eta\delta
M^2_\eta(\mu)\nonumber\ .
\ena
The parameter $B$ is related to the vacuum condensate $B_0$ :
\eq
B=B_0+2(m_s+2{\widehat m})Z_0^S(\mu)\ .\lbl{B}
\en
The constants $Z_0^S$ and $Z_0^P$ are defined in terms of the
low energy behaviour
of the two point functions $<S^0S^0-S^8S^8>$ and $<P^0P^0-P^8P^8>$,
respectively.
They both violate the Zweig rule : $Z_0^S$ is expected to be small (compared to
$A_0$) since it is suppressed in the large $N_c$-limit. $Z_0^P$, on
the other hand,
receives a contribution from the axial anomaly and, in principle, does not need
to be
small. Both $Z_0^P$ and $Z_0^S$ are order parameters of chiral symmetry
breaking. Finally, $\delta M^2_\pi,\ \delta M^2_K$ and $\delta M^2_\eta$
contain
chiral logarithms, which compensate for the scale dependences
 of $A_0$ and $Z_0^S$, and
higher powers of quark masses. A few comments are in order:

\indent

{\bf i) The overall convergence of the expansions} \rf{exp} is controlled
by the small parameter
\eq
{m_q\over \Lambda_H}\ ,\ \Lambda_H \sim 1\hbox{GeV}\ .\lbl{mq}
\en
This statement becomes particularly transparent in the large $N_c$ limit, in
which the chiral logarithms drop out. In general, the coefficient of the n-th
power of quark masses is defined by means of the low momentum behaviour of
a n-point
function of quark bilinears \rf{sp}. (An example is the coefficient $A_0$, c.f.
Eq. \rf{Pisp}). These n-point functions satisfy superconvergent dispersion
relations (they are order parameters) which can be saturated by
the lowest massive
bound states of mass $\sim \Lambda_H\sim 1$ GeV. This leads to the estimate
that
the coefficient of $m_q^n\ (n\geq 2)$ in \rf{exp} should indeed be of the order
$\Lambda_H^{2-n}$ times a factor of order unity. Notice that quark
confinement makes
it difficult to extend the above estimate to the quark condensate $B_0$ itself.

{\bf ii) The relative importance of the linear term} in \rf{exp} is
controled by another parameter,
\eq
{m_q\over m_0}\ ,\quad m_0={B_0\over 2A_0}\ ,\lbl{m0}
\en
which indicates how small the quark masses should actually be such as to
ensure the validity of the Gell-Mann--Oakes--Renner formula. If
$m_s\sim m_0$, the
contributions of the first and second order terms in the expansion of
$M_K^2$ and of $M^2_\eta$
would become comparable, and if even $m_u,\ m_d\sim m_0$, the same would
happen for the pion mass. There is no compelling reason why $m_0$ should be
as large
as $\Lambda_H$ and there is nothing unnatural in
having $m_q\sim m_0\ll \Lambda_H$.
As already pointed out, $B_0$ (at the 1 GeV scale) might be as small as 100 MeV
$\sim F_0$, and $A_0$ can be estimated using the superconvergent dispersion
relations for the two-point function \rf{A0} : the standard QCD
sum-rule technique
leads to an order of magnitude estimate
\eq
A_0\vert_{\hbox{1 GeV}}\simeq 1\div 5\ .\lbl{A0est}
\en
(Notice that for small $B_0$, the $\mu$-dependence of $A_0$ is very weak).
Hence, the antiferromagnetic alternative for the QCD vacuum is perfectly
consistent with a low value of $m_0=(10\div 50)$ MeV,
indicating the possibility of a
strong violation of the Gell-Mann--Oakes--Renner formula, for $M^2_\eta,\
M^2_K$
and even for $M^2_\pi$.

{\bf iii) The Gell-Mann--Okubo mass formula} is compatible with Eq. \rf{exp}
 even if the first order term does not dominate. From \rf{exp}, one obtains
\eq
3F^2_\eta M^2_\eta+F^2_\pi M^2_\pi-4F^2_KM^2_K=4(m_s-\widehat m)^2F^2_0
(A_0+2Z_0^P)+\cdots\ ,\lbl{gmo}
\en
where the ellipsis represents higher order contributions arising from $\delta
M^2_P$. Neglecting the latter (and the splitting of the decay constants $F_P$),
the GMO formula can even become exact, provided that $A_0+2Z_0^P=0$. Notice,
however,
that a priori there is no reason for this relation to hold. Actually it can
hardly
hold exactly, since it is badly violated for $N_c\to\infty$: in this limit
$A_0=O(1)$, whereas $Z_0^P=O(N_c)$, reflecting the vanishing of the $\eta'$
mass
in the chiral limit as $N_c\to\infty$ \cite{Ven}. The fact
that the GMO relation is well
satisfied by experimental masses does not by itself imply the dominance of the
quark condensate term in the expansion of Goldstone boson masses.

{\bf iv)} The parameter which actually controls the relative size of the first
and second order terms in Eqs. \rf{exp} is {\bf the quark mass ratio}
\eq
r={m_s\over\widehat m}={2m_s\over m_u+m_d}\ .\lbl{r}
\en
It is well known that provided the first term dominates, $r$ is given by the
ratio of $M^2_K$ and $M^2_\pi$, $r\simeq$26 \cite{GOR,W1,GL2}. To
the extent that $B_0$
decreases, the
ratio $r$ becomes smaller, until it reaches a critical value corresponding to
vanishing $B_0$. For $r$ below this critical value, the vacuum would become
unstable with respect to small perturbations produced by quark masses.

Eqs. \rf{exp} put the above statements under a quantitative control. From
the first two equations, one obtains
$$\eqalignno{
{{2\widehat mB_0}\over{M_\pi^2}}&={1\over r^2-1}[(r-r_1^*)(r+r_1^*+2)-
2(r^*_2-r)(r+2)\zeta]\left({{M_\pi^* F_\pi}\over{M_\pi F_0}}\right)^2&(12a)\cr
4{{\widehat m^2A_0}\over{M_\pi^2}}&=2\,{r_2^*-r\over r^2-1}\,
\left({{M_\pi^* F_\pi}\over{M_\pi F_0}}\right)^2\ .&(12b)\cr
}$$
\setcounter{equation}{12}
Here,
\bea
(M^*_\pi)^2&=&M^2_\pi-\delta M^2_\pi\ ,\quad (M^*_K)^2=M^2_K-\delta M^2_K\ ,
\nonumber\\
& &\lbl{rcrit}\\
r_1^*&=&2\,{F_KM^*_K\over F_\pi M^*_\pi}-1\ ,\quad r^*_2=2\left({F_KM_K^*\over
F_\pi M^*_\pi}\right)^2-1\ ,\nonumber
\ena
and $\zeta$ stands for (a small) Zweig rule violating parameter,
\eq
\zeta={Z_0^S\over A_0}\ .\lbl{zeta}
\en
The QCD vacuum is likely to be in the phase in which the order
parameters $m_q B_0,\ A_0$
and $Z_0^S$ are all non-negative. This phase is characterized by
\eq
r_1^*\leq r\leq r_2^*\ ,\ \
0\leq \zeta\leq {1\over 2}\ {r-r_1^*\over r_2^*-r}\ {r+r_1^*+2\over r+2}\
.\lbl{bounds}
\en
For $r=r_2^*$, the order parameter $A_0$ vanishes and the pion mass is given by
the vacuum condensate $B_0 : r_2^*$ may be referred to as the {\bf
ferromagnetic
critical point}. On the other hand, for $r=r_1^*$, the quark condensate $B_0$
vanishes and the pion mass is entirely accounted for by the order parameter
$A_0$: $r_1^*$ is the {\bf anti-ferromagnetic critical point}. The values of
$r_1^*$ and $r_2^*$ may be expanded in powers of quark masses,
\eq
r_1^*=r_1+O(m_q)\ ,\quad r_2^*=r_2+O(m_q)\ ,
\en
where the leading order values $r_1$ and $r_2$ are known :
\eq
r_1=2\,{M_K\over M_\pi}-1\simeq 6.3\ ,\quad r_2=2\,{M^2_K\over M^2_\pi}-1=
25.9\ .
\en
The corrections push both $r_1^*$ and $r_2^*$ upwards: $r_1^*$ can
reach the value
$8\div 9$, whereas $r_2^*$ can actually be as large as 38 (this can be seen
upon neglecting, in Eqs. \rf{rcrit}, $\delta M^2_\pi$ and $\delta M^2_K$, and
by using the experimental value $F_K/F_\pi$=1.22).

The question, whether the value of the quark mass ratio $r$ is closer to the
ferromagnetic or to the antiferromagnetic critical point, has to be decided
experimentally.

\section {\bf Expansion of the effective Lagrangian}

\indent
The above remarks do not affect the construction of the low energy effective
Lagrangian \cite{W2}. Its form is merely dictated by the chiral symmetry
\cite{Leut} and
by the transformation properties of the symmetry breaking quark mass term, and
there is obviously no question to alter these fundamental properties of QCD.
${\cal L}^{eff}$ is a function of 8 Goldstone boson fields
(conventionally collected into an SU(3) element U) and of external sources
${ v}_\mu,\ { a}_\mu,\ { s}$ and ${ p}$, the scalar source
${s}$ containing the quark mass matrix ${\cal M}$.
The formalism and notation used here are standard \cite{GL1}, unless
otherwise stated. ${\cal L}^{eff}$ consists of an infinite tower of
invariants
\eq
{\cal L}^{eff}=\sum_{(k,l)}{\cal L}_{(k,l)}\ ,\lbl{Leff}
\en
where ${\cal L}_{(k,l)}$ contains $k$ powers of covariant derivatives and
$l$ powers of
scalar or pseudoscalar sources. In the low energy limit,  ${\cal L}_{(k,l)}$
vanishes
like the $k$-th power of external momenta $p$ and the $l$-th power of
the quark mass $m_q$,
\eq
{\cal L}_{(k,l)}\sim p^k m_q^l\ .
\en
Chiral perturbation theory
is an expansion of ${\cal L}^{eff}$ in powers of the pion (kaon)
mass assuming that all external momenta are of that size. For sufficiently
small quark masses,
such that both $m_q\ll\Lambda_H$ {\bf and} $m_q\ll m_0={B_0\over 2A_0}$ hold,
one has
${\cal L}_{(k,l)}=O(p^{k+2l})$. In this case, one can write
\eq
{\cal L}^{eff}={\cal L}^{(2)}+{\cal L}^{(4)}+{\cal L}^{(6)}+\cdots\lbl{scpt}
\en
where \cite{GL1}
\eq
{\cal L}^{(d)}=\sum_{k+2l=d}{\cal L}_{(k,l)}\ .
\en
This expansion defines the standard $\chi$PT. If, on the other hand, for actual
values of quark masses one has $m_q\sim m_0={B_0\over 2A_0}\ll\Lambda_H$, both
$m_q$ and $B_0$ should count as parameters of the size of the pion mass and,
consequently, ${\cal L}_{(k,l)}=O(p^{k+l})$. This new counting yields
a  different
expansion of ${\cal L}^{eff}$,
\eq
{\cal L}^{eff}=\tilde{\cal L}^{(2)}+\tilde{\cal L}^{(3)}
+\tilde{\cal L}^{(4)}+\tilde{\cal L}^{(5)}+\tilde{\cal
L}^{(6)}+\cdots\lbl{gcpt}
\en
where \cite{FSS}
\eq
\tilde{\cal L}^{(d)}=\sum_{k+l+n=d}B_0^n{\cal L}_{(k,l)}\ .\lbl{gexp}
\en
It should be stressed that Eqs. \rf{scpt} and \rf{gcpt} represent
two different expansions
of the {\it same} effective Lagrangian.
To all orders they are identical, at a given
finite order they may differ.

It is straightforward to write down the most general expression \cite{FSS} of
${\tilde{\cal L}}^{(2)}$ which defines the leading $O(p^2)$ order of G$\chi$PT
:
\bea
{\tilde{\cal L}}^{(2)}&=&{1\over 4}F_0^2
\left\{\langle D_\mu U^+D^\mu U\rangle +2B_0\langle U^+\chi+\chi^+U\rangle+
\right.\nonumber\\
\qquad &+&A_0\langle (U^+\chi)^2+(\chi^+U)^2\rangle
+Z_0^S<U^+\chi+\chi^+U\rangle ^2\lbl{glead}\\
\qquad &+&Z_0^P\langle U^+\chi-\chi^+U\rangle ^2
+\left. H_0\langle \chi^+\chi\rangle\right\}\ .\nonumber
\ena
Here $\chi$ collects the scalar and pseudoscalar sources,
\eq
\chi={s}+i{ p}={\cal M}+\cdots \ ,\, {\cal M}={\rm diag}(m_u,m_d,m_s)\ .
\en
Notice the absence of the factor $2B_0$, which appears in the standard
definition of $\chi$ \cite{GL1}. The meaning
of this difference will shortly become
obvious. Compared to ${\cal L}^{(2)}$, Eq. \rf{glead} contains
additional terms. The
constants $A_0,\, Z_0^S$ and $Z_0^P$ are the same as introduced in the previous
section : Eq. (5), (with $F_\pi=F_K=F_\eta=F_0$ and $\delta M^2_P=0$)
is indeed a
straightforward consequence of the Lagrangian \rf{glead}.
The fact that the $A_0,\
Z_0^S$ and $Z_0^P$ terms now appear at the same order $O(p^2)$ as the $B_0$
term
reflects the possibility that in Eqs. (5) the first order and second order
terms
are of comparable size. To this order $O(p^2)$, the low energy constants in
\rf{glead}
can be expressed in terms of physical masses $M_\pi,\ M_K,\ M_\eta$, of
the quark mass
ratio $r=m_s/\widehat m$ and of the Zweig rule violating parameter $\zeta$
\rf{zeta}.
Expanding Eqs. (12a), (12b) and \rf{gmo}, one gets
$$\eqalignno{
{2\widehat mB_0\over M^2_\pi}&={1\over r^2-1}
[(r-r_1)(r+r_1+2)-2(r_2-r)(r+2)\zeta]&(26a)\cr
{4\widehat m^2A_0\over M^2_\pi}&=2\,{r_2-r\over r^2-1}&(26b)\cr
{4\widehat m^2Z^S_0\over M^2_\pi}&=2\,{r_2-r\over r^2-1}\zeta&(26c)\cr
{4\widehat m^2Z^P_0\over M^2_\pi}&={1\over (r-1)^2}\ {\Delta_{GMO}\over
2}-{r_2-r\over r^2-1}\ ,&(26d)\cr
}$$
\setcounter{equation}{26}
\noindent
where $\Delta_{GMO}\equiv(3M^2_\eta-4M_K^2+M^2_\pi)/M_\pi^2$. It is seen
that for the
particular values of the parameters $r=r_2,\ \Delta_{GMO}=0$, Eqs. (26) imply
$A_0=Z_0^S=Z_0^P=0$ and $2\widehat mB_0=M^2_\pi$, i.e. one recovers the
standard
$O(p^2)$ Lagrangian ${\cal L}^{(2)}$. Order by order, G$\chi$PT contains the
standard $\chi$PT as a special case.

The next to the leading term of the expansion \rf{gcpt} is
of odd order $O(p^3)$, which is a new
feature, absent in the standard $\chi$PT. One has \cite{KMS}
\bea
\tilde{\cal L}^{(3)}&=&{1\over 4}F_0^2
\left\{\xi\langle D_\mu U^+D^\mu U(\chi^+U+U^+\chi)\rangle
+\tilde\xi\langle D_\mu U^+D^\mu U\rangle
\langle\chi^+U+U^+\chi\rangle\right.\nonumber\\
\quad&+&\rho_1\langle (\chi^+U)^3+(U^+\chi)^3\rangle
+\rho_2\langle (\chi^+U+U^+\chi)\chi^+\chi\rangle\nonumber\\
\quad&+&\rho_3\langle\chi^+U-U^+\chi\rangle
\langle(\chi^+U)^2-(U^+\chi)^2\rangle\lbl{L3}\\
\quad&+&\rho_4\langle(\chi^+U)^2
+(U^+\chi)^2\rangle
\langle\chi^+U+U^+\chi\rangle\nonumber\\
\quad&+&\rho_5\langle\chi^+\chi\rangle\langle\chi^+U+U^+\chi\rangle\nonumber\\
\quad&+&\left.\rho_6\langle\chi^+U-U^+\chi\rangle^2
\langle\chi^+U+U^+\chi\rangle+\rho_7\langle\chi^+U+U^+\chi\rangle^3
\right\}\ .\nonumber
\ena
In writing down Eq. \rf{L3}, we have adopted some
conventions which are worth to be
specified. The parameters of $\tilde{\cal L}^{(2)}+\tilde{\cal L}^{(3)}$ are
finite, (divergences only start at order $O(p^4)$) and they may be viewed as
independent variables. It is convenient to tag each of these variables by a
QCD correlation function. An example is the $O(p^2)$ parameter $A_0$, closely
related to the two-point function \rf{A0}. This relation
may be further specified as
\eq
A_0(\mu)=A_0+O(B_0^2\ln\mu)\ ,\lbl{A0exp}
\en
where $A_0(\mu)$ is {\sl defined} by Eq. \rf{Pisp}. Here, the statement
is that the
expansion of $A_0(\mu)$ in powers of $B_0$ does not contain a linear term. This
constrains the way one writes $\tilde{\cal L}^{(3)}$ :

{\bf i)} A term $B_0{\cal L}_{(0,2)}$ which, according to Eq. \rf{gexp} could
be
present in $\tilde{\cal L}^{(3)}$, is obviously irrelevant. It can be absorbed
into the covariant transformation of sources $\chi$ and of parameters of ${\cal
L}^{eff}$ which has been discovered some times ago by Kaplan and Manohar
\cite{KaMa},
(see also Ref. \cite{DW}). Indeed, Eq. \rf{A0exp} may be
viewed as a physical condition
fixing the reparametrization ambiguity of ${\cal L}^{eff}$ pointed out in
\cite{KaMa} and in \cite{DW}. In general, the $B_0$
dependent terms in Eq. \rf{gexp}
will be
introduced only if they are required by renormalization. This is not the case
of $\tilde{\cal L}^{(3)}$.

{\bf ii)} Similarly, a ${\cal L}_{(2,1)}$ term like
$\langle D_\mu U^+D^\mu\chi+D_\mu\chi^+D^\mu U\rangle$
which would yield a contribution to Eq. \rf{A0exp} linear in $B_0$, can be
transformed away by a source dependent redefinition of Goldstone boson fields.
Notice that this convention has not been used in  Ref. \cite{FSS}.

The main physical effect described by the Lagrangian $\tilde{\cal L}^{(3)}$
is the splitting of the decay constants $F_\pi,\ F_K$, and $F_\eta$.
One easily finds
\bea
{F_\pi^2\over F_0^2}&=&1+2\widehat m\xi+2(2\widehat m+m_s)\tilde\xi\nonumber\\
{F_K^2\over F_0^2}&=&1+(m_s+\widehat m)\xi+2(2\widehat m+m_s)\tilde\xi
\lbl{F}\\
{F_\eta^2\over F_0^2}&=&1+{2\over 3}(\widehat m+2 m_s)\xi+2(2\widehat
m+m_s)\tilde\xi\ .\nonumber
\ena
This allows to express the constant $\widehat m\xi$ as
\eq
\widehat m\xi={1\over r-1}
\left({F_K^2\over F_\pi^2}-1\right)\lbl{mxi}\ ,
\en
whereas the Zweig rule violating parameter $\tilde\xi$ remains at this stage
undetermined. The ${\cal L}_{(0,3)}$ part of $\tilde{\cal L}^{(3)}$,
described by
the constants $\rho_1\cdots \rho_7$ generates an $O(m^3_q/\Lambda_H)$
contribution
to the pseudoscalar masses. Notice that in the standard $\chi$PT these terms
would count as $O(p^6)$.

$\tilde{\cal L}^{(3)}$ provides the simplest example of odd chiral orders
characteristic of G$\chi$PT. They do not correspond to an increase in the
number of loops, but to additional corrections in powers of the quark masses.
In
the standard $\chi$PT \cite{GL1}, the splitting in
the decay constants is a $O(M^2_\pi)$
effect arising from loops (tadpoles) and from the corresponding counterterms
contained in ${\cal L}^{(4)}$. Here, the leading contribution \rf{mxi} counts
as $O(M_\pi)$, (actually, it can hardly be expressed in terms of the pion
mass), and the loop effects only show up at the next, $O(p^4)$, order. Notice
that for $r\simeq 10$, the constant \rf{mxi} is of the order
$\widehat m\xi=(5\div
6)\%$ - a typical size of other $O(M_\pi)$ effects, such as the deviation from
the Goldberger-Treiman relation.

The Lagrangian $\tilde{\cal L}^{(4)}$ describing the next order, $O(p^4)$,
consists of several components :
\eq
\tilde{\cal L}^{(4)}={\cal L}_{(4,0)}+{\cal L}_{(2,2)}+{\cal L}_{(0,4)}+
B_0^2{\cal L}_{(0,2)}^\prime+B_0{\cal L}_{(2,1)}^\prime+B_0{\cal L}_{(0,3)}^
\prime\ .\lbl{L4}
\en
${\cal L}_{(4,0)}$ is the part of the standard ${\cal L}^{(4)}$, which
consists of four derivatives and contains no $\chi$, i.e. no quark mass :
${\cal L}_{(4,0)}$ is given by the standard five terms \cite{GL1}
described by the low energy
constants $L_1,\ L_2,\ L_3,\ L_9$ and $L_{10}$. ${\cal L}_{(2,2)}$
is a new term,
which in the standard $\chi$PT would count as $O(p^6)$, whereas
${\cal L}_{(0,4)}$ involves 4 insertions of a quark mass, and in the standard
$\chi$PT it would be relegated up to the order $O(p^8)$. ${\cal L}_{(2,2)}$ and
${\cal L}_{(0,4)}$ involve about 20 independent terms each. An experimental
determination of all the corresponding low energy constants is obviously hard
to imagine. However, a few particular combinations of these constants which
contribute, for instance, to $K_{l3}$ and $K_{l4}$ decays, or to the $\pi\pi$
scattering amplitude at the one loop level, can be estimated and included into
the analysis.

The last three terms in Eq. \rf{L4} represent $B_0$-dependent
counterterms of order
$O(p^4)$ which are needed to renormalize one loop divergences that arise from
using the vertices of $\tilde{\cal L}^{(2)}$ alone in the loop. They
renormalize
the constants $A_0,\ Z_0^S$ by higher order contributions, of order $O(B_0^2)$,
and the constants $\xi,\ {\tilde\xi}$ and $\rho_i$ by an amount $O(B_0)$. In
generalized $\chi$PT, renormalization proceeds order by order in the expansion
in powers of the constant $B_0$.

The standard $O(p^4)$ order Lagrangian ${\cal L}^{(4)}$ contains 10
low energy constants $L_i$ \cite{GL1}. They are all involved in
$\tilde{\cal L}^{(2)}+\tilde{\cal L}^{(3)}+\tilde{\cal L}^{(4)}$. As already
pointed out, $L_1,\ L_2,\ L_3,\ L_9$ and $L_{10}$ - the constants of
${\cal L}_{(4,0)}$ - play an identical role in both schemes. $L_4$ and $L_5$
are
related to the constants $\tilde\xi$ and $\xi$ in $\tilde{\cal L}^{(3)}$ :
\bea
L_4^r&=&{F_0^2\over 8B_0}
\left\{\tilde\xi+O\left({B_0\over {(4\pi F_0)^2}}\ln\mu\right)\right\}\ ,
\nonumber\\
& &\lbl{Lia}\\
L_5^r&=&{F_0^2\over 8B_0}
\left\{\xi+O\left({B_0\over {(4\pi F_0)^2}}\ln\mu\right)\right\}\nonumber\ .
\ena
Finally, the standard $O(p^4)$ constants $L_6,\ L_7,\ L_8$ are  related to the
 $\tilde{\cal L}^{(2)}$ constants $Z_0^S,\ Z_0^P$ and $A_0$, respectively :
\bea
L_6^r&=&{F_0^2\over 16B_0^2}
\left\{Z_0^S+O\left({B_0^2\over {(4\pi F_0)^2}}\ln\mu\right)\right\}\ ,
\nonumber\\
L_7^r&=&{F_0^2\over 16B_0^2}Z_0^P\ ,\lbl{Lib}\\
L_8^r&=&{F_0^2\over 16B_0^2}
\left\{A_0+O\left({B_0^2\over {(4\pi F_0)^2}}\ln\mu\right)\right\}\ .\nonumber
\ena

On the other hand, ${\tilde{\cal L}}^{(2)} + {\tilde{\cal L}}^{(3)} +
{\tilde{\cal L}}^{(4)}$ involves additional terms, not contained in ${\cal
L}^{(2)} + {\cal L}^{(4)}$, and which the standard $\chi$PT relegates to higher
orders $d>4$. Setting the corresponding additional constants to zero, one
recovers the standard $\chi$PT up to and including order $O(p^4)$. This
phenomenon is general: Order by order, the standard expansion reappears as a
special case of G$\chi$PT.

\section{\bf Examples of differences between standard
and $\quad$ generalized $\chi$PT}

\indent
The difference between standard and generalized $\chi$PT merely concerns the
symmetry breaking sector of the theory. In the chiral limit $m_u=m_d=m_s=0$,
observables described by purely derivative terms of the effective Lagrangian
${\cal L}_{(2,0)}+{\cal L}_{(4,0)}+{\cal L}_{(6,0)}+\cdots$
are essentially given by
soft pion theorems which are identical in both schemes. This concerns, in
particular, the electromagnetic radius of the pion, the decay $\pi\to
e\nu\gamma$, the $\pi\pi$ scattering in the P-wave, the leading order $K_{l3}$
and $K_{e4}$ form-factors, etc...

\noindent The difficulty of disentangling both schemes resides in the fact that
symmetry breaking effects are small and not well known experimentally. In this
section we are going to give an overview of the main differences
between the predictions
of standard and generalized $\chi$PT for symmetry breaking effects which
already appear at the leading order $O(p^2)$. Those have more chances to be
observable, at least indirectly.

{\bf 4a - $m_d-m_u$ : Dismiss of the ellipse} \cite{Ell}

We have already stressed the importance of the quark mass ratio
$r={m_s/\widehat m}$ as an independent parameter of G$\chi$PT. When isospin
breaking is switched on, G$\chi$PT leads to a relation between the two quark
mass ratios $r$ and $R$, with
\eq
R={m_s-\widehat m\over m_d-m_u}\ .\lbl{R}
\en
Taking a suitable linear combination of the expansions of $M^2_{\pi^+},\
M^2_{K^0}$ and $M^2_{K^+}$, the $O(m_{q})$ terms, the $O(m^2_{q})$
terms and even the $O(m^2_{q}\ln m_q)$ terms drop out and one obtains
\cite{FSS}
\eq
R\Delta M^2_K-(M^2_K-M^2_\pi)-{1\over 2}\ {r-1\over r+1}\ (r_2-r)M^2_\pi=O
\left({m^3_q\over \Lambda_H}\right)\ ,\lbl{massrel}
\en
where $\Delta M^2_K=M^2_{K^0}-M^2_{K^+}$ is the kaon mass difference in the
absence of electromagnetism. Eq. (35) holds independently of the value of $r$,
i.e. both in the standard and in the generalized $\chi$PT. In
the standard case, Eq. \rf{massrel}
should be rather accurate, whereas in the generalized setting the
neglected $O(m^3_q)$ terms - arising from ${\cal L}_{(0,3)}$ and given by the
low energy constants $\rho_i$ \rf{L3} - can easily represent a 30\% correction.

To the best of our knowledge, Eq. \rf{massrel} has
never appeared in the literature on standard
$\chi$PT. Instead, one rather expands {\sl ratios} of pseudoscalar meson masses
and one eliminates the $O(m_q)$ terms from this expansion (see e.g. Eqs.
(10.11)
and (10.17) of the second of Refs. \cite{GL1}). In this
way, one arrives at the well known
elliptic relation between ${m_s\over m_d}$ and ${m_u\over m_d}$, extensively
commented upon in the literature \cite{Ell}. The difference between
the two procedures
is instructive : When expanding the {\sl linear
combinations} of pseudoscalar meson
masses, one expands in powers of the small parameter
${m_q\over\Lambda_H}$. On
the other hand, when expanding {\sl mass ratios}, one assumes, in addition,
that the
parameter  ${m_{q}\over m_0},\ (m_0={B_0\over 2A_0})$ is also small.
Actually, the elliptic relation \cite{Ell} should read
\eq
{1\over Q^2}\left({m_s\over m_d}\right)^2+
\left({m_u\over m_d}\right)^2=1+O\left[\left({m_s\over m_0}\right)^2\right]\ ,
\lbl{ellipse}
\en
where
\eq
{1\over Q^2}={M^2_\pi\Delta M^2_K\over M^2_K(M^2_K-M^2_\pi)}\ .
\en
The quark mass ratios lie on the ellipse only provided $m_s\ll m_0$ or,
equivalently, provided $r={m_s\over\widehat m}$ is close to $r_2=25.9$, whereas
Eq. \rf{massrel} has a more general validity.

It is interesting to look at Eq. \rf{massrel} in the light
of recent discussions of a
possible large violation of Dashen's theorem \cite{Dash}. We can write
\eq
\Delta M^2_K=(M^2_{K^0}-M^2_{K^+})_{exp}+\gamma(M^2_{\pi^+}-M^2_{\pi^0})\ ,
\lbl{dash}
\en
where the parameter $\gamma$ describes the departure from Dashen's theorem
: $\gamma=1$ if the theorem is exact. Various recent estimates \cite{Dash}
expect
$\gamma$ somewhere between 1 and 2. Taking in Eq. \rf{massrel}
$r=25.9$, one obtains
$R=43$ for $\gamma=1$ and $R=34.8$ for $\gamma=2$. For $r=10$, Eq. \rf{massrel}
predicts $R=66$ if $\gamma=1$, whereas $R=53$ if $\gamma=2$. Assuming $R\sim45$
(as expected from baryon masses and from $\omega-\rho$ mixing \cite{GL2})
we observe that
by increasing $\gamma$, we are left with less room for $O({m_s\over
\Lambda_H})$
corrections to Eq. \rf{massrel}, which could provide a
valuable information on the $\tilde{\cal L}^{(3)}$ parameters $\rho_i$.

$$ $$

{\bf 4b - Quark condensates for $m_q\ne 0$}

The parameter $B_0$ describes the quark condensate in the chiral limit
$m_u=m_d=m_s=0$. The leading order Lagrangian $\tilde{\cal L}^{(2)}$ allows to
express $<\Omega\,\vert\bar qq\vert\,\Omega>$, where $\vert\,\Omega>$
is the ground
state for $m_q\ne 0$, beyond this limit. One gets
\eq
-{F_0^{-2}}<\Omega\,\vert\bar qq\vert\,\Omega>=B_0+m_q(A_0+{1\over 2}H_0)+
4(\widehat m+{1\over 2}m_s)Z_0^S+O({{m^2_q}\over{\Lambda_H}})\ .\lbl{condexp}
\en
where $q=u,d,s$ denotes a given quark flavour. The point is that if
$B_0\ll\Lambda_H$, the $O(m_q)$ contribution is Eq. \rf{condexp} can
be relatively
important. Comparing for instance the term $m_sA_0$ to $B_0$ one finds, using
Eqs. (26a) and (26b), $m_sA_0\sim 2.3 B_0$, assuming $r=10$ and $\zeta=0$. This
should not be surprising : in G$\chi$PT {\sl all} contributions of
$\tilde{\cal L}^{(2)}$ are supposed to be of the same order of magnitude. In
practice, it implies that the $m_q\ne 0$ condensates can exhibit a large
flavour dependence. Unfortunately, there is no way to pin down the constant
$H_0$ which controls this dependence quantitatively : $H_0$ is not a low energy
order parameter, but rather a short distance counterterm (see Ref. \cite{GL1}
for a discussion of this point).

{\bf 4c - Large corrections to the soft pion theorems}

Within generalized $\chi$PT the corrections to soft pion theorems can
sometimes be rather important - of the same order as the soft pion result
itself. Consider, for instance, the scalar form factor of the pion at
vanishing momentum transfer or, equivalently, the pion $\sigma$-term
\eq
\delta^{ij}\sigma_\pi(0)=<p,\pi^i\vert\widehat m(\bar uu+\bar dd)\vert p,\pi^j>
\en
The soft pion result for $\sigma_\pi(0)$ is well known :
\eq
\sigma_\pi(0)\vert_{\hbox{soft pions}}=2\widehat mB_0\ .\lbl{sigpi}
\en
Since in the G$\chi$PT, $2\widehat mB_0$ can be considerably {\sl smaller}
than $M^2_\pi$, one might be tempted to conclude, on the basis of the
soft pion result \rf{sigpi}, that in G$\chi$PT $\sigma_\pi(0)$ is
smaller than in
the standard theory. This conclusion does not take into account {\sl all}
$O(p^2)$ contributions to  $\sigma_\pi(0)$ described by $\tilde{\cal L}^{(2)}$.
Writing $\sigma_\pi(0)=\widehat m{\partial\over\partial\widehat m}M^2_\pi$,
using Eqs. \rf{exp} and (26), and neglecting the Zweig rule violating parameter
$\zeta$, one obtains the correct $O(p^2)$ result,
\eq
\sigma_\pi(0)=\widehat m{\partial M^2_\pi\over\partial\widehat m}=
\left(1+2\,{r_2-r\over r^2-1}\right)M^2_\pi\ .\lbl{sigp2}
\en
It is seen that, when $r$ decreases from $r=r_2$ to $r=r_1\simeq 6.3$, the
pion $\sigma$-term {\sl increases} from $M^2_\pi$ to $2M^2_\pi$. In G$\chi$PT,
the soft pion result will receive a large correction, whenever the soft pion
theorem result is proportional to the quark condensate $B_0$. The
reason is that both $B_0$
and $m_q$ count as quantities of order $O(M_\pi)$. The formalism of
G$\chi$PT automatically takes care of such large corrections. This phenomenon
can in principle lead to modifications of the standard evaluation of the
non-leptonic K-decay matrix elements in the large $N_c$ limit, in particular,
of the penguin-contribution to the ratio
${\epsilon^\prime\over\epsilon}$ \cite{EdR}.

There is another relevant example of a similar nature : The $\chi$PT
prediction for the low $Q^2$ behavior of the spectral function associated
with the divergence of the axial current,
\eq
\rho(Q^2)={1\over 2\pi}
\sum_n(2\pi)^4\delta^4(Q-P_n)
\left\vert<n\vert\partial^\mu(\bar d\gamma_\mu\gamma_5
u)\vert0>\right\vert^2\ .\lbl{rho}
\en
At low $Q^2$, the continuum part of $\rho(Q^2)$ is dominated by the
contribution of $3\pi$ intermediate states. Using for the latter the soft pion
theorem (and neglecting in the phase space integral \rf{rho} the pion mass) one
gets the standard result \cite{DdR}
\eq
\rho_{3\pi}(Q^2)={1\over 768\pi^4}\ {M^4_\pi\over F^2_\pi}\
Q^2+\cdots\lbl{rho3st}
\en
Within the $O(p^2)$ G$\chi$PT, this result is considerably modified. Still
neglecting the pion mass in the phase space integral (in order to facilitate a
comparison with the standard result) one gets \cite{SFK}
\eq
\rho_{3\pi}(Q^2)={1\over 768\pi^4}\ {M^4_\pi\over F^2_\pi}\ Q^2
\left\{1+10\,{r_2-r\over r^2-1}+30\left({r_2-r\over
r^2-1}\right)^2\right\}+\cdots\lbl{rho3g}
\en
As $r$ decreases from $r_2$ down to $r_1$, the enhancement factor in Eq.
\rf{rho3g}
increases from 1 to 13.5. This enhancement would considerably affect the
existing estimates of $\widehat m$ using the QCD sum rules \cite{DdR}.

{\bf 4d - $\pi\pi$ and $\pi K$ scattering}

Our last example of a leading order difference between the standard and the
generalized $\chi$PT is of a direct experimental relevance: It concerns the low
energy $\pi\pi$ \cite{FSS} and $\pi K$ \cite{KSSF} scattering. The $\pi\pi$
amplitude
predicted by the leading order G$\chi$PT lagrangian ${\tilde{\cal L}}^{(2)}$
reads
\eq
A_{lead}(s\vert t,\,u)= {1\over{F_0^2}}\,(s-2{\widehat m}{\tilde B})\
,\lbl{Wein}
\en
where
\eq
{\tilde B}= B_0 + 2m_sZ_O^S
\en
is the non strange quark antiquark condensate in the SU(2)$\times$SU(2) chiral
limit (see Eq. \rf{condexp}). Eq. \rf{Wein} holds both in the standard case and
in the genertalized $\chi$PT. In the standard case, however, $2{\widehat
m}{\tilde B}\sim M_\pi^2$, and one recovers the well known formula first
obtained by Weinberg \cite{W3}. In G$\chi$PT, this formula is modified
already at
the leading order $O(p^2 )$: $2{\widehat m}{\tilde B}$ can be considerably
smaller than $M_\pi^2$ by an amount which depends on $r$. For $r$ decreasing
from $r_2$=25.9 to $r$=6.3, $2{\widehat m}{\tilde B}$ decreases from $M_\pi^2$
to zero. The low energy $\pi\pi$ scattering thus provides us with a
quasi-unique experimental access to the order parameter ${\widehat m}B_0$. The
corresponding G$\chi$PT predictions, endowed with the necessary loop
corrections, are presented in detail in the section on $\pi\pi$ interactions of
the present Handbook.

A similar conclusion holds in the case of $\pi K$ scattering \cite{KSSF}. The
latter gives
a contribution to the $K_{l4}$ form factor $R$, which is measurable in the
$K_{\mu 4}$ decay mode. Whereas the leading order predictions for the $K_{l4}$
form factors $F$, $G$ and $H$ are identical in both schemes, there is a
detectable difference in the leading order expression for $R$ \cite{KSSF}.
Whether
this difference can be observed in practice is presently under investigation
\cite{KSl4}.

\section{\bf Values of the low energy constants}

\indent
It is not incorrect to state that G$\chi$PT is more general just because it
admits a considerably wider range of values of certain low energy constants and
of the current quark masses than the standard scheme. On the other hand, the
standard $\chi$PT claims that the values of the $O(p^4)$ constants $L_1$,...,
$L_{10}$ are well under control, both determinig them from data \cite{GL1}
and estimating
them via resonance saturation \cite{Eck,EckL}. Similar claims are
often made about the
values of the light quark masses \cite{PDG}. The standard values of the low
energy parameters can indeed be justified within the set of assumptions
underlying the standard $\chi$PT. However, beyond this framework, the same
experimental data, the same sum rules, etc..., can often yield rather different
results. We start by discussing the constants $L_i$ from the point of view of
G$\chi$PT.

{\bf 5a - $L_{10}$ and $L_9$}

These two constants appear in ${\cal L}_{(4,0)}$, which is the part
common to both ${\cal L}^{(4)}$ and ${\tilde{\cal L}}^{(4)}$. Their
measurements
via the electromagnetic radius of the pion and the radiative decay $\pi\to
e\nu\gamma$ \cite{GL1} should not be altered
in G$\chi$PT. Furthermore, $L_{10}$ and $L_9$
are, respectively, related to two and three point functions of vector and axial
vector currents dominated by vector and axial vector meson poles
\cite{EckL,DH}. This makes
the estimates of $L_9$ and of $L_{10}$ rather stable.

{\bf 5b - $L_3$, $L_2$ and $L_1$}

These remaining three constants of ${\cal L}_{(4,0)}$ are more difficult to
measure. Although they are not directly related to explicit symmetry breaking
effects, they enter the observables ($K_{l4}$ form factors \cite{Kl4}, $\pi\pi$
D-waves \cite{GL1}) together with $r$-dependent loop corrections. Consequently,
for lower values of $r$, the standard determination of $L_1^r$, $L_2^r$ and
$L_3$ is slightly modified \cite{KSl4}, by not more than a factor of two.

The estimates of these constants via resonance saturation is also more involved
since, in addition to vector and axial vector mesons, $L_1$, $L_2$ and $L_3$
receive a contribution from scalar exchanges, which are not known so well
\cite{Eck}. The
reason of this complication is the fact that, unlike $L_9$ and $L_{10}$, $L_1$,
$L_2$ and $L_3$ are related to four point functions of the vector and axial
currents.

{\bf 5c - $L_4$ and $L_5$}

Let us concentrate on the ${\cal L}_{(2,1)}$ constant $L_5$ ($L_4$ violates the
Zweig rule) and, for simplicity, let us stick to the leading large $N_c$
behaviour, denoting the leading part of $L_5$ by ${\hat L}_5$. One has (Eq.
(32)),
\eq
{\hat L}_5 = {{\xi F_\pi^2}\over{8B_0}} \sim {{F_K^2 -
F_\pi^2}\over{8(m_s-{\widehat m})B_0}}\ ,
\en
where we have used (30). The standard determination of ${\hat L}_5$ \cite{GL1}
(or, equivalently, of $L_5^r$ including
the chiral logarithms) would replace $(m_s -
{\widehat m})B_0$ by $M_K^2 - M_\pi^2 $, leading to the value ${\hat L}_5\sim
2.2\times 10^{-3}$. Within G$\chi$PT, this last step could be misleading,
provided $r$ is well below $r_2\sim 25.9$. Using instead the leading order
formula (26a) and neglecting the Zweig rule violating parameter $\zeta$, one
obtains
\eq
{\hat L}_5 = {{F_K^2 -
F_\pi^2}\over{4M_\pi^2}}\,{1\over{r-r_1}}\,{{r+1}\over{r+r_1+2}}\ .
\en
Hence, the value of ${\hat L}_5$ one extracts from Eq. (48) crucially depends
on the quark mass ratio $r$: For $r=r_2\sim25.9$ one gets the ``standard
value'' ${\hat L_5}\sim 2.2\times 10^{-3}$, whereas for $r$ decreasing down to
$r_1$, ${\hat L}_5$ increases up to infinity. The estimates of ${\hat L}_5$ via
resonances merely concern scalar exchanges, whose description is considerably
more ambiguous \cite{MS} than in the case of vector and axial vector mesons
\cite{EckL}.

{\bf 5d - $L_6$, $L_7$ and $L_8$}

The values of these ${\cal L}_{(0,2)}$ constants are at the heart of our
discussion of symmetry breaking effects. $L_6^r$ violates the Zweig rule in the
$0^{++}$ channel and it will not be discussed here. The combination $L_8^r +
2L_7$ can be related to the deviation ${\Delta}_{GMO}$ from the
Gell-Mann--Okubo mass formula \cite{GL1} (see Eqs. (10) and (32)). The standard
evaluation of $L_8^r +2L_7$ based on this relation suffers from a similar bias
as in the case of ${\hat L}_5^r$: The expression for $L_8^r + 2L_7$ involves
the
factor $(m_s -{\widehat m})^2B_0^2$ in the denominator. As a result, the value
of $L_8^r +2L_7$ is even more sensitive to $r$ than ${\hat L}_5$, and it can
actually be considerably larger than the standard value \cite{GL1} $L_8^r +2L_7
\sim 0.1\times 10^{-3}$. A separate measurement of $L_8$ is usually based on
the relation with the isospin breaking quark mass ratio $R$ (34), which is
known from different sources \cite{GL1,GL2,Ell}. In G$\chi$PT, the relation
between $L_8$ and $R$
can easily break down, for the same reason which could invalidate the
elliptic relation
(36): The importance of unduly neglected $O(m_s/m_0)$ contributions. As a
consequence, the standard value \cite{GL1} of $L_8^r(M_\rho)=(0.9\pm 0.3)\times
10^{-3}$ can be underestimated by as much as two orders of magnitude. Writing
\eq
{\hat L}_8 ={{F_0^2}\over{16M_0^2}}\ ,\, M_0^2=2m_0B_0={{B_0^2}\over{A_0}}\ ,
\en
the renormalization group invariant mass parameter $M_0$ can be estimated from
Eqs. (26a,b). In the standard scheme ($r\sim r_2$), $M_0$ is expected to be at
a
GeV scale, whereas in G$\chi$PT, $M_0$ can be as small as $M_\pi$, or even
smaller.

The estimate of ${\hat L}_8$ from resonance contributions can be obtained from
the two point function (3) which satisfies the superconvergent sum rule
\cite{MS}
\eq
\int\,dq^2 \Im m \Pi_{SP}(q^2) = 0\ .
\en
Saturating this sum rule by nothing but the pion and a single $0^{++}$ state of
mass $M_S$,
one obtains $M_0 = M_S\sim 1$ GeV, leading to a value for ${\hat L}_8$ which is
of the standard order of magnitude $\sim 10^{-3}$. Unfortunately, the above
argument does not by itself support the standard picture, and can be
turned around: If
$B_0\ll\Lambda_H$, the pion contribution ($=B_0^2$) can hardly dominate the
pseudoscalar component of the sum rule (51). In order to balance the scalar
contribution, one has to include an excited $\pi^{\prime}$ state of mass
$M_P>M_S$, $M_P\simeq (1200\div 1300)$MeV. It is then easy to see that the mass
parameter $M_0$ in Eq. (50) can take any value between zero and $M_S$
\cite{MS}.

Quite generally, the introduction of $J=0$ resonances, compatible with chiral
symmetry and with the short distance properties of QCD correlation functions,
into the effective
lagrangian \cite{Eck} is more tricky and more ambiguous
than in the case of $J=1$ states \cite{EckL}. The authors of Ref. \cite{Eck}
have, for
instance, decided to disregard the contribution of the $\pi^{\prime}$ nonet.
Doing so, they have a priori eliminated the low $B_0$ alternative.

The standard $\chi$PT rewrites the expansion in quark masses as an expansion in
powers of $M_\pi^2\sim2{\widehat m}B_0$. In the vicinity of the
antiferromagnetic critical point, $r\sim r^*_1$, the coefficients of the latter
expansion, viz. $L_4$...$L_8$, blow up, and the expansion has to be redefined.
G$\chi$PT is precisely such a redefinition. At the leading order, it is
characterized by a single undetermined parameter not present in the standard
scheme, the quark mass ratio $r= m_s/{\widehat m}$. One may say that G$\chi$PT
parametrizes the deviations from standard predictions of $\chi$PT in terms of
the deviation of $r$ from its standard value $r^{st}\sim r_2$ \cite{GL1}.

{\bf 5e - Running light quark masses}

In G$\chi$PT, the decrease of the quark mass ratio $r$ from $r\sim 25$ to
$r\sim 10$ is likely to be interpreted as an increase in the value of the
running quark mass ${\widehat m}(\mu)$ by a factor of $2\div 3$. The order of
magnitude of the mass differences $m_s -{\widehat m}$ and $m_d - m_u$ should
remain essentially unchanged. In this connection, it should be stressed that
all existing estimates of ${\widehat m}(\mu)$ use in one way or the other the
assumption $r\sim 25$. This is obvious for those approaches which deduce the
value of ${\widehat m}(\mu)$ from the estimates of $m_s - {\widehat m}$
\cite{W1,GL2}. It is however even true in the case of direct quantitative
determinations of ${\widehat m}(\mu)$ from QCD sum rules \cite{DdR}. The latter
express the square of ${\widehat m}(\mu)$ as a weighted integral of the
spectral function (43). Nothing is known experimentally about $\rho (Q^2)$
beyond the one pion intermediate state contribution. The only existing attempt
to fill this gap makes use of the low $Q^2$ behaviour (44) of
$\rho_{3\pi}(Q^2)$ as given by the
standard $\chi$PT in order to normalize the whole ${\rho}_{3\pi}(Q^2)$
contribution
represented by a broad $\pi^{\prime}$ Breit-Wigner peak. In this way, the
result ${\widehat m}(1{\rm GeV})=(7\pm 1){\rm MeV}$ is obtained \cite{DdR}.
If, instead,
the $3\pi$ contribution is normalized using the G$\chi$PT result (45), the
above value of ${\widehat m}(\mu)$ is increased by a factor $2\div 3$. (Let us
note in passing that within G$\chi$PT, the possibility of having $m_u = 0$
appears even  less likely than in the standard case \cite{Ell}.)

In general, G$\chi$PT admits and expects a larger absolute strength of the
divergence of the $\Delta S=0$ axial current away from the pion pole than the
standard scheme could possibly tolerate. The physical origin of this increase
would be ascribed to the importance of the ``$\pi^{\prime}$ contribution'',
which can and should be checked experimentally: The ${\rho}_{3\pi}(Q^2)$
component of the spectral function (43) can be measured in high statistics
$\tau\to 3\pi\nu_\tau$ decay experiments \cite{SFK}, and in this way, the
determination of ${\widehat m}(\mu)$ could be put on a solid experimental
basis.

Concluding this section, it is worth emphasizing that nothing in the preceeding
discussion indicates that the standard determination of the $L_i$'s and of the
light quark masses is internally inconsistent. The standard $\chi$PT together
with the standard value of the low energy parameters is a perfectly
self-consistent scheme. However, it is not the only possible consistent
scheme and it does not contain its proper justification.

\section{Experimental tests}

\indent
The framework of G$\chi$PT can be exploited in order to measure the quark mass
ratio $r=m_s/{\widehat m}$ together with a few other low energy constants in
several independent experiments. Since for $r=r^{st}\sim 25.9$, the predictions
of the standard $\chi$PT are always contained as a special case, measurements
of $r$ constitute a powerfull test of the assumptions underlying the standard
scheme ($B_0\sim\Lambda_H$). In order to control all relevant low energy
parameters which are needed up to a given degree of precision, a simultaneous
analysis of many processes may be necessary. In this way, one may also hope to
achieve some control of higher orders of $\chi$PT not included into the
analysis. At present, there are a few experimental issues which appear to be
particularly relevant.

{\bf 6a - Deviations from the Goldberger-Treiman relation} \cite{GT}

The comparison of the quark mass expansion of the deviations from the
Goldberger-Treiman relations in the three channels $NP$, $\Lambda P$, $\Sigma
N$, yields a first order measurement of $m_s/{\widehat m}$. The result is very
sensitive to the precise values of the strong coupling constants $g_{\pi NN}$,
$g_{K\Lambda N}$ and $g_{K\Sigma N}$. Present values \cite{GPNN} lead
to the bound $r\leq 10.6\pm 4.2 $, which requires a confirmation. Da$\Phi$ne
could contribute to a new
precise determination of the hyperon coupling constants from low energy $KN$
and ${\bar K}N$ scattering data.

{\bf 6b - Low energy $\pi\pi$ scattering}

Standard $\chi$PT leads to firm predictions \cite{GL1} for scattering lengths
and low energy phase shifts. If $r$ decreases, G$\chi$PT reveals that the
$\pi\pi$ interaction in the I=0 S-wave becomes stronger \cite{FSS}, in
agreement with existing (but not very accurate) data. The detailed discussion
may be found in the $\pi\pi$ Section of the present Handbook.

Da$\Phi$ne can accurately measure the low energy phase shifts $\delta_0^0 -
\delta_1^1$ as a function of energy. This information, taken together with the
direct determination of $a_0^0 -a_0^2$ to 5\% accuracy from the $\pi^+\pi^-$
atom lifetime experiment which is planed at CERN \cite{CZA}, could well provide
a decisive measurement of $m_s/{\widehat m}$ and a crucial test of $\chi$PT.

{\bf 6c - $K_{l4}$ form factors} \cite{Kl4,KSl4}

The $\pi\pi$ phases and the $K_{l4}$ form factors are closely tied together. A
simultaneous analysis is necessary in order to determine the constants $L_1$,
$L_2$ and $L_3$ and the Zweig rule violating parameters $\zeta$ and
${\tilde\xi}/\xi$, and to reduce the theoretical uncertainty in the predictions
for the $\pi\pi$ scattering amplitude. In addition, the form factor $R$
measurable in the $K_{\mu4}$ decay mode could provide an indirect access to the
$K\pi$ scattering amplitude \cite{KSSF,KSl4}.

{\bf 6d - $\gamma\gamma\to\pi^0\pi^0$}

The existing low energy data \cite{Mars} are not very accurate. They can be
reproduced by G$\chi$PT to the one loop, provided $r\leq 10$ \cite{KMS}.
Unfortunately, the stringent disagreement of the one loop result of standard
$\chi$PT does not yet allow to conclude in favour of smaller values of $r$: At
the two loop level, the standard $\chi$PT restores the agreement with the data
without problems \cite{BGS}. More accurate data in the threshold region could
help to clarify this situation.

{\bf 6e - Azimuthal asymmetries in the decay $\tau\to 3\pi\nu_\tau$} \cite
{SFK}

The measurement of these angular asymmetries at a per cent level would provide
a direct information about the magnitude of the divergence of the axial current
in the $\pi^\prime$ region. When combined with QCD sum rules \cite{DdR}, this
measurement yields a probe of the value of the running quark mass ${\widehat
m}(\mu )$ \cite{SFK}. This test of G$\chi$PT has the virtue of going beyond its
own framework. Unfortunately, it also goes beyond the scope of Da$\Phi$ne,
requiring a tau-charm factory or the like.

The above list is not exhaustive. Additional $K$, $\eta$ or $\eta^\prime$
decays might be of interest.

\indent

\indent

\end{document}